\documentclass[%
reprint,
 amsmath,amssymb,
 aps,
longbibliography]{revtex4-2}

\usepackage{graphicx}
\usepackage{dcolumn}
\usepackage{bm}
\usepackage{comment}
\usepackage{xcolor}
\usepackage[colorinlistoftodos]{todonotes}
\usepackage{hyperref}
\usepackage{mathtools}
\usepackage[english]{babel}
\usepackage{subfigure}

\begin{document}


\title{Non-resonant Density of States Enhancement at Low Energies for Three or Four Neutrons}

\author{Michael D. Higgins}
\email{higgin45@purdue.edu}
\affiliation{Department of Physics and Astronomy, Purdue University, West Lafayette, Indiana 47907 USA}
\author{Chris H. Greene}
\email{chgreene@purdue.edu}
\affiliation{Department of Physics and Astronomy, Purdue University, West Lafayette, Indiana 47907 USA}
\affiliation{Purdue Quantum Science and Engineering Institute, Purdue University, West Lafayette, Indiana 47907 USA}
\author{A. Kievsky}%
 \email{alejandro.kievsky@pi.infn.it}
 \author{M. Viviani}
  \email{michele.viviani@pi.infn.it}
\affiliation{%
 Instituto Nazionale di Fisica Nucleare, Largo Pontecorvo 3, 56127 Pisa, Italy
}%


\date{\today}

\begin{abstract}
The low energy systems of three or four neutrons are treated within the adiabatic hyperspherical framework, yielding an understanding of the low energy quantum states in terms of an adiabatic potential energy curve.  The dominant low energy potential curve for each system, computed here using widely accepted nucleon-nucleon interactions with and without the inclusion of a three-nucleon force, shows no sign of a low energy resonance.  However, both systems exhibit a low energy enhancement of the density of states, or of the Wigner-Smith time-delay, which derives from long-range universal physics analogous to the Efimov effect.  That enhancement could be relevant to understanding the low energy excess of correlated 4-neutron ejection events observed experimentally in a nuclear reaction by Kisamori et al.\cite{Kisamori} 
\end{abstract}

\maketitle



The three- and four-neutron (3n and 4n) systems are intriguing and important problems in few-nucleon fundamental physics that deserve a comprehensive, deep theoretical understanding.  While no 4n bound state is generally believed to exist, there have been speculations for decades about the possible existence of a long-lived resonance in the 4-particle scattering continuum.  Those early speculations have evolved into renewed interest triggered by the recent experimental observation of an enhanced signal of 4 low energy neutrons emerging together, which they tentatively interpreted as a possible 4n resonance (or bound) state, by Kisamori {\it et al.}\cite{Kisamori}. The present Letter investigates the possible existence of a low energy resonance-like enhancement of the density of states in both the 4n and 3n systems, using well established nucleon-nucleon (NN) interactions, with and without the inclusion of a three-nucleon force (3NF), and also using a simple Gaussian potential adjusted to match the neutron-neutron (nn) scattering length and effective range.  

In our study, the low energy regions of the 3n and 4n systems are explored using the adiabatic hyperspherical representation, which has a strong track record of successfully predicting and interpreting resonances for atomic systems.\cite{lin1975PRL,botero1986PRL}  Our results with the aforementioned potentials are consistent with strong enhancements of the low-energy density of states (or Wigner-Smith time delay)  for both the 3n and 4n systems, although the nature of the potential curves and the eigenphaseshift energy dependences make it clear that the enhanced density of states should not be viewed as a resonance.  Moreover, neither the 3n nor the 4n system is close to possessing a bound state.  Our analysis also demonstrates how the density of states enhancement can be understood in terms of universal physics considerations that are closely related to the Efimov effect.\cite{DIncaoReview,Rittenhouse-2011JPB,Greene2017RMP}

Remarkably, theoretical treatments to date have not been able to reach a consensus agreement about whether a 3n or 4n resonance exists, consistent with the presently understood NN interaction potentials. The need for more theoretical input into this problem is therefore clear, given the conflicting conclusions reached so far by competing theoretical methods.
Specifically, some of the theory published to date is consistent with the claimed experimental observation of a low energy resonance in the 4n system,\cite{ShirokovVary2016prl,Gandolfi2017prc,LiMichelHu2019PRC} whereas alternative theoretical analyses are incompatible with a resonance or bound state interpretation of the experimental measurement \cite{HiyamaLazauskas2016PRC,Fossez,Deltuva4n2018PL,
DeltuvaLazauskas2019PRC,DeltuvaLazauskas2019commentPRL,GandolfiHammer2019PRL,HiyamaKamimura2018FPhys}.  An advantage of the present method based on the adiabatic hyperspherical representation is that the absence of a resonance state is immediately clear visually after inspecting the relevant adiabatic potential energy curve for the system.  Moreover, our quantitative calculation shows that a nonresonant density of states enhancement is guaranteed to be present at low energies, owing to the attractive hyperradial potential energy at very long range.  Specifically, this connects with the universal behavior of three- and four-fermion systems close to the unitarity limit.  We propose that such a density of states enhancement could help to understand the enhanced production of  four low energy neutrons in the experiment of Kisamori {\it et al.}\cite{Kisamori}, even in the absence of a tetraneutron resonance state.

The theoretical approach adopted here starts 
considering realistic nuclear interaction Hamiltonians. 
They are constructed by an overall fit of the existing $np$ and $pp$ data and, invoking 
charge symmetry invariance, they can be applied to describe neutron systems as well.
In particular, we have considered the AV18 
and AV8' NN potentials~\cite{Wiringa:1994w} as well as the recent local NN potentials
derived within the chiral effective field theory approach~\cite{Piarulli:2016vel,Baroni:2018fdn}, in particular the model NV2-Ia. 
With the AV18 potential, we have performed calculations with the inclusion of the Urbana 
and Illinois 3NFs.\cite{UIX,UIX2,Illinois3BF} It should be noticed that
the two-body singlet $nn$ scattering length is large and negative, believed to be approximately 
$a \approx -18 {\rm fm}$, consistently reproduced by the NN interactions considered.  
Motivated by the large value of the $nn$ scattering length, 
we have also carried out calculations using a simple single Gaussian potential, adjusted to describe that 
value and the corresponding effective range, in order to explore connections with universal behavior and 
the unitary limit of the three- and four-fermion systems. The form of the Gaussian potential used is $V(r)=V_{0}\mathrm{exp}(-r^2/r_{0}^2)$, where $V_{0}$ is the strength of the potential and $r_0$ is the range. The parameters used for singlet and triplet interactions are given in Table \ref{table:gauss_params}.
\begin{table}[ht]
\caption{Simple Gaussian parameters used for singlet and triplet two--body interactions. The parameters were extracted from fits to the central component of the $\mathrm{AV8}^{\prime}$ potential.} 
\centering
\label{table:gauss_params}
\resizebox{0.5\columnwidth}{!}{\begin{tabular}{l|l|l}
    $\mathrm{State}$ & $V_{0} (\mathrm{MeV})$ & $r_{0} (\mathrm{fm})$\\
    \hline
    $\prescript{1}{}{S}$ & $-31.7674$ & $1.7801$\\
    $\prescript{3}{}{P}$ & $95.7280$ & $0.8809$ \\\hline 
\end{tabular}}
\end{table}
In all our calculations it has been found that 
the use of a particular form of NN potential, with or without the inclusion of the 3NF, has comparatively 
little influence on the results; in particular the inclusion of 3NFs only slightly modified the potential
 curves around 1 - 2 fm, making them more repulsive.
 The 3n and 4n Schr\"odinger equations 
are then solved in the adiabatic hyperspherical representation~\cite{macek1968JPB,lin1975PRL,fano1983RPPb,Garrido2014prc}, 
which has a proven track record in correctly predicting resonances, especially in atomic and molecular 
physics contexts. After one diagonalizes the fixed-hyperradius Hamiltonian, $H_{\rho=const}$, the 
$\rho$-dependent eigenvalues $U_\nu(\rho)$ act as adiabatic potential energy curves (and couplings 
$W_{\nu,\nu'}$) that often make it immediately and visibly clear whether or not there is a resonance, and 
they yield an immediate interpretation if a resonance does exist~\cite{DIncaoReview}. Note for reference 
that two successful predictions and interpretations of atomic shape resonances, carried out within the 
adiabatic hyperspherical framework in Refs.\cite{lin1975PRL,botero1986PRL}, were eventually confirmed by
both experiment\cite{bryant1977PRL, MichishioNagashima2016NatureComm} and by other theory for the singlet 
electronic $L^\pi=1^-$ states of the negative ions H$^-$ and Ps$^-$.

The greatest numerical challenge in the present study is the calculation of the 4n- and 3n- potential energy curves $U_\nu(\rho)$ and the elements of the coupling matrix operator \textcolor{black}{$W_{\nu,\nu'}(\rho)=-\frac{\hbar^2}{2\mu}{(} \langle \Phi_\nu | \frac{\partial}{\partial\rho} \Phi_{\nu'} \rangle\frac{\partial}{\partial\rho} + \langle \Phi_\nu |\frac{\partial^2}{\partial\rho^2} \Phi_{\nu'}\rangle  {)} $ }, where $\Phi_\nu$ are the adiabatic eigenfunctions.
Our approach tackles this variationally at each value of $\rho$, by expanding the unknown adiabatic eigenfunctions ($\Phi_\nu$) into a basis set.  Two different choices of the basis set have been implemented in our study.  The first is a set of coupled hyperspherical harmonics and spinors adapted to the symmetry of interest, e.g. $J^\pi=0^+$ for the tetraneutron.  The second type of basis set implemented to solve the fixed-$\rho$ Schr\"odinger equation is a linear combination of correlated Gaussian functions.\cite{stecher2009PRA,RakshitBlume2012pra,Daily2014pra,Suzuki2020prc}  Following diagonalization of $H_{\rho=const}$ at each $\rho$, a Rayleigh-Ritz {\it upper bound} on the exact potential $U_\nu(\rho)$ is obtained. The following theorem is important for our subsequent analysis below:  {\emph{When the hyperradial Schr\"odinger equation is solved in the lowest potential curve, including also just the diagonal nonadiabatic coupling terms $W_{\nu,\nu}(\rho)$, the lowest computed energy of the system will be a rigorous upper bound to the exact ground state energy.}}  Much of our detailed analysis of the resonance physics has been performed in the adiabatic approximation, which neglects off-diagonal coupling terms. Our tests show its general validity for these 3n and 4n systems.

To understand the basic idea of the formulation, consider first the one-dimensional hyperradial Schr\"odinger equation. The single adiabatic term variational ansatz for the wavefunction is written for $N$ particles in their relative frame as:  $\Psi(\rho,\Omega)=\rho^{-(3N-4)/2}\Phi_0(\rho;\Omega)F_0(\rho)$, where $\Phi_0(\rho;\Omega)$ is the lowest adiabatic eigenfunction of $H_{\rho=const}$ with eigenvalue $U_0(\rho)$ and repulsive diagonal correction term $W_{00}(\rho)$.  The radial equation then takes the form:
\begin{equation}
-\frac{\hbar^2}{2\mu} \frac{d^2}{d\rho^2} F_0(\rho) + (u_0(\rho)-E)F_0(\rho)=0,    
\end{equation}
where the full, effective adiabatic potential in the lowest channel, including the diagonal correction term, is:
\begin{equation}
u_0(\rho)\equiv U_0(\rho) + W_{00}(\rho).
\end{equation}
Note that $u_0(\rho)$ includes the effective centrifugal term $\frac{\hbar^2}{2\mu} \frac{(3N-6)(3N-4)}{4\rho^2}$ associated with the elimination of first order hyperradial derivatives from the effective radial Schr\"odinger equation.
Here $\mu$ is \textcolor{black}{a reference mass (we use $\mu=m/2$ with $m$ the neutron mass)}, and the hyperradius $\rho$, for equal mass particles is defined
by $ \rho^2 \equiv \frac{2}{N} \Sigma_{i<j} r_{ij}^2$,
where $r_{ij}$ is the distance between neutrons $i$ and $j$.  

It is known from universality studies that for $N$-particle systems dominated by a large magnitude two-body scattering length $a$, their lowest long range hyperradial potential energy curve in the continuum has the following asymptotic form, at $\rho\rightarrow \infty$:
\begin{equation}
\label{Eq:AsymForm}
u_0(\rho) \rightarrow \frac{\hbar^2}{2\mu} {\biggr (}\frac{l_{\rm eff}(l_{\rm eff}+1)}{\rho^2}
+ C\frac{a}{\rho^3} {\biggr )},
\end{equation}
where $C$ and $l_{\rm eff}$ depend on the number of particles and their statistics; their values are given in Table \ref{table:1s2sscattlengths} below for the symmetries considered in the present study. The adiabatic correction term $W_{00}(\rho)$ decays asymptotically at least as fast as $\rho^{-4}$ for the 3n and 4n systems and therefore has no role in the above decomposition. For the present problem, where the $nn$ scattering length is large and negative, the attractive long range term proportional to $a/\rho^3$ has key implications for the low energy Wigner-Smith time-delay\cite{Wigner1955PR,Smith1960PR}, $Q=2 \hbar d \delta/dE$, which also measures the density of states of the system.\cite{OrangeReview}  In particular the density of states diverges like $E^{-1/2}$ as $E\rightarrow 0$ since the scattering phaseshift $\delta(E)$ at low energy can be seen  perturbatively  to equal $\delta \rightarrow - C a k/(2{l_{\rm eff}}+2{l_{\rm eff}}^2)$ as the wavenumber $k \rightarrow 0$. 

\begin{table}[ht]
\caption{Unitarity (subscript $u$) and non--unitarity (no subscript) long--range ($\rho \rightarrow \infty$) coefficients of the lowest adiabatic potential (see Eq.(\ref{Eq:AsymForm})). Our values of $l_{\rm eff}$ extracted at unitarity are shown (a), as are the corresponding values at unitarity obtained in accurate calculations by Yin and Blume(b).\cite{YinBlume2015pra}} 
\centering
\label{table:1s2sscattlengths}
\resizebox{1\columnwidth}{!}{\begin{tabular}{l|l|l|l|l|l}
    $N$ & $(LS)J^{\pi}$ & $l_{\rm eff}$ & $C$ & $l_{\rm eff,u}^{(a)}$ & $l_{\rm eff,u}^{(b)}$\\ 
    \hline
    $3$ & $(1\frac{1}{2})\frac{3}{2}^-$ & $5/2$ & $15.22$ & $1.275$ & $1.2727(1)$\\
    $4$ & $(00)0^+$ & $5$ & $86.68 $ & $2.027$ & $2.0091(4)$ \\\hline 
\end{tabular}}
\end{table}

Next consider the numerical computation of the adiabatic hyperspherical potential energy curves for the 3n and 4n systems.
We use two different variational basis sets, an expansion into hyperspherical harmonics (extremely accurate at small and intermediate values of $\rho$)\cite{KievskyViviani2008,KievskyViviani2020FoP,Daily-2015} and an expansion into correlated Gaussian basis functions (more accurate at large $\rho$)\cite{stecher2009PRA,RakshitBlume2012pra,Daily2014pra}. The HH basis produces well converged results for the quantities of interest,
$U_\nu(\rho)$ and $W_{\nu,\nu}(\rho)$, in a relatively large range of $\rho$ values, $0-50{\rm fm}$ and
$0-30{\rm fm}$ for 3n and 4n respectively. At the end of this region all potential models considered almost collapse onto a single adiabatic curve and therefore one particular model can be used for calculating the adiabatic curves beyond that point. To this purpose we have used the correlated Gaussian hyperspherical basis set (CGHS)~\cite{stecher2009PRA,Daily2014pra} in connection with the AV8' interaction, which has a Gaussian expansion that efficiently connects with the CGHS method\cite{HiyamaPrivate}.


The lowest adiabatic hyperspherical potential energy curves in the most attractive symmetries of the 4n and 3n systems, namely $0^+$ and ${\frac{3}{2}}^-$ respectively, are plotted in Fig.1.  At a glance it is immediately apparent that the lowest potential curve for both systems is totally repulsive, and moreover positive at all hyperradii, which guarantees both that there is no bound state and that there can be no resonance state in the low energy range below 10 MeV.  Nevertheless there is extensive attraction in the system, which is apparent from the fact that the potential curve lies everywhere well below the upper dashed curve which would apply if there were zero interaction between the neutrons.  Over much of the range of $\rho$, in fact, both systems are slightly closer to the unitary-limiting potentials that would emerge if the two-body potential was made even more attractive to give an infinite singlet n-n scattering length (i.e., closer to the lower dashed curves in Fig.1 ), than to the noninteracting limit.

\begin{figure} 
\begin{frame}{}
\hspace{-0.0in}\includegraphics[width=8.5 cm] {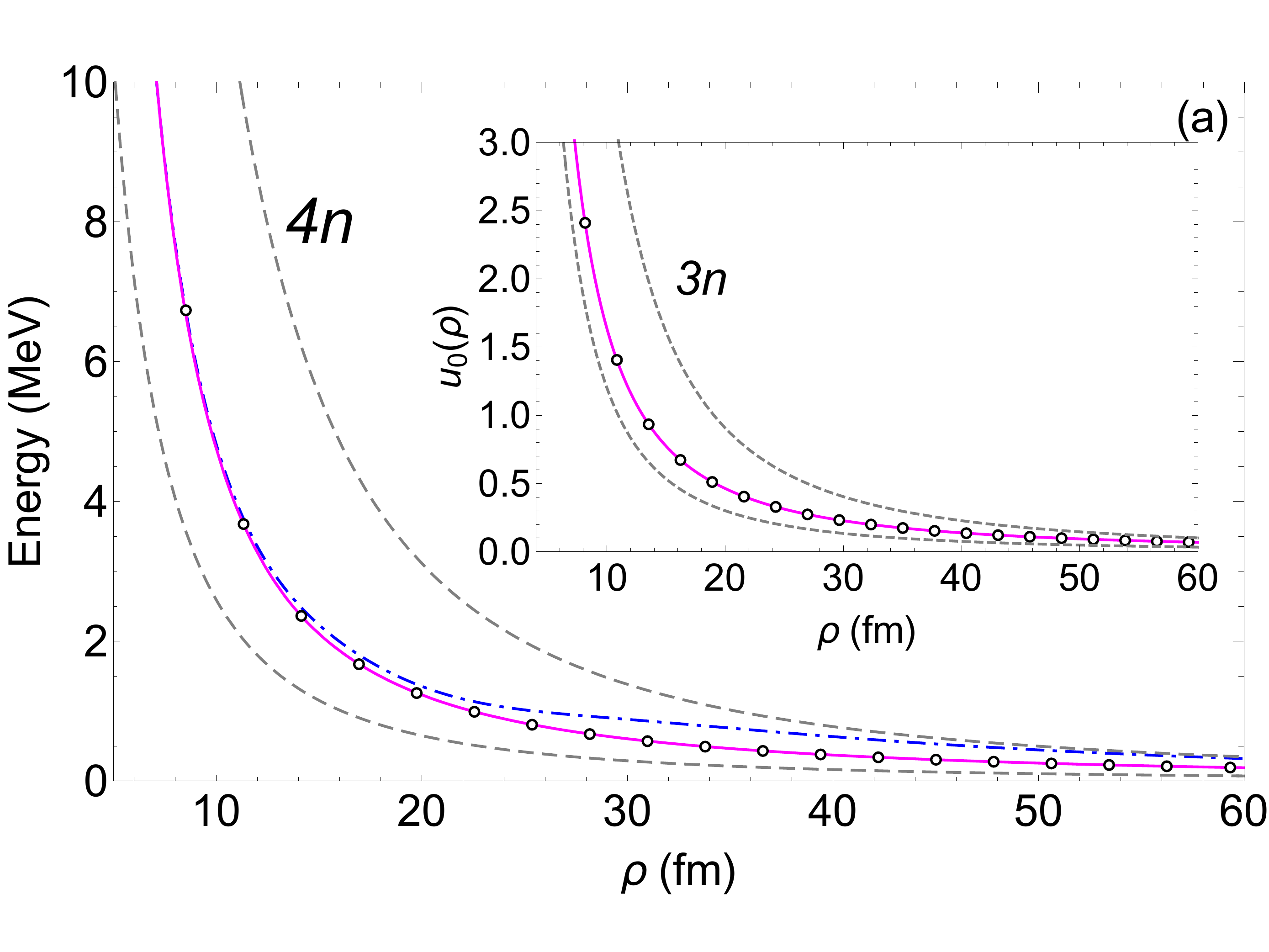}
\end{frame}
\vskip-0.3in 
\begin{frame}{}
\hspace{-0.0in}\includegraphics[width=8.5 cm] {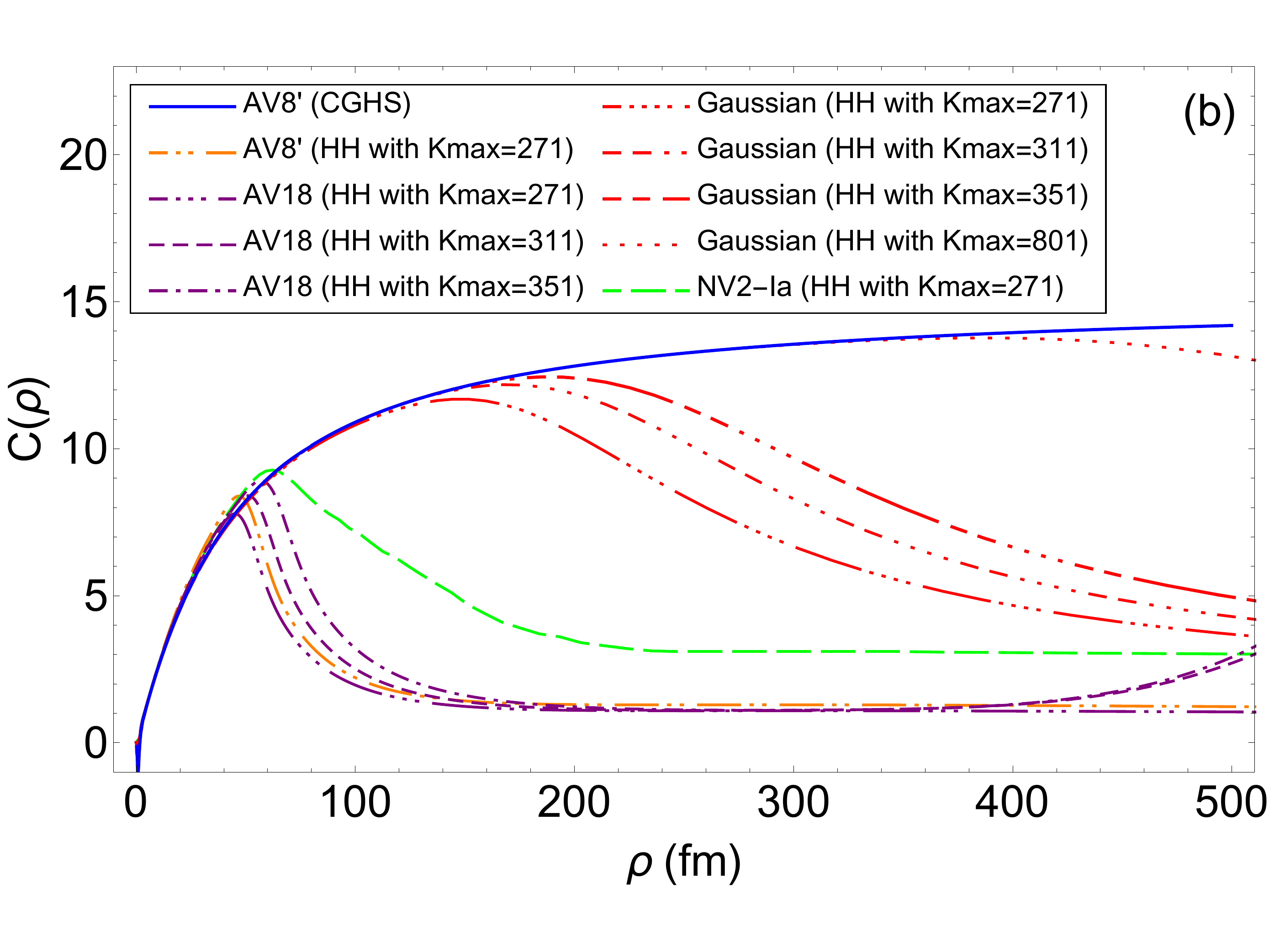}
\end{frame}
\vskip-0.15in 
\caption{\label{fig:Fig1}  (a) Hyperspherical potential curve for the most attractive channel ($0^+$ for 4n,  $\frac{3}{2}^-$ for 3n in the inset) for both the four-neutron and three-neutron systems.  Comparison of the lowest $0^+$ 4n adiabatic hyperspherical potential energy curves computed with the HH method \textcolor{black}{(blue dash--dotted curve, which shows the best calculation for the AV18 Hamiltonian with Kmax=140 and is not accurately converged at hyperradii beyond approximately 20 fm)}. The lowest solid magenta 4n (and 3n in the inset) potential energy curve is computed using the \textcolor{black}{CGHS method applied to the AV8' Hamiltonian}.
    The \textcolor{black}{open circles} are the adiabatic potentials calculated using a simple NN Gaussian potential (see text).
    The lower dashed gray curves in both the main figure and the inset are the expected long range $\rho^{-2}$ potentials at unitarity for this symmetry of the 4n and 3n systems, i.e. in the infinite scattering length limit (see text and Table \ref{table:1s2sscattlengths}).  The upper dashed gray curves are the corresponding potentials for noninteracting neutrons.  Clearly there is no local minimum and no local maximum of the type that is always associated with a quasi-bound resonance. (b) Plot of the function $C(\rho) \equiv (\rho/a) [\rho^2 u_0(\rho) 2 \mu/\hbar^2 - l_{\rm eff}(l_{\rm eff}+1) ]$ for the 3n case. According to Eq.(3), we should obtain $C(\rho \rightarrow \infty)=C$, where $C$ is the coefficient listed in Table \ref{table:1s2sscattlengths}. We observe the slow convergence for large $\rho$ of the adiabatic potentials calculated using the HH basis. However, it has to be noted that where the convergence is achieved, the functions $C(\rho)$ obtained for the different  interactions used in this work almost collapse onto a single curve. Noticeably, this happens already for fairly small values of $\rho$, showing that the adiabatic potentials are already universal at moderate values of the hyperradius. In fact,  the limit $C(\rho)=C$ is reached only for $\rho>500\ {\rm fm}$. }
\end{figure}
The HH expansion includes the eigenfunctions of the grand angular momentum operator 
${\bf K^2}$, with eigenvalues $K(K+7)$ (4n) and $K(K+4)$ (3n), with values of $K\ge 2$ up to a maximum 
value $K^{max}=140$ (4n) and $K^{max}=801$ (3n). \textcolor{black}{In the CGHS expansion, only natural parity states are treated with 208 $L$=0 and 92 $L$=2 basis functions for the 4n system and 57 $L$=1 basis functions for the 3n system.} 
The potentials $u_0(\rho)$ in Fig.~\ref{fig:Fig1} include the repulsive diagonal correction term 
$W_{00}(\rho)$ and, in the 4n case, is shown for \textcolor{black}{the largest value} of $K^{max}$ \textcolor{black}{(represented by the dash--dotted curve).}
The upper dashed curve is the expected asymptotic form of the lowest noninteracting potential curve, 
namely $u^{NI}_0(\rho) \rightarrow \frac{30\hbar^2}{2\mu \rho^2}$ for the 4n $0^+$ symmetry.  
The lower dashed curve is the effective potential at unitarity, i.e. 
$u^{univ}_0(\rho) = \frac{{l_{\rm eff,u}}({l_{\rm eff,u}}+1) \hbar^2}{2\mu \rho^2}$, with ${l_{\rm eff,u}}$ given in 
Table~\ref{table:1s2sscattlengths} for both the 3n and 4n systems; these values would result if the 
neutrons interacted through a zero-range potential that produces an infinite singlet $nn$ scattering 
length\cite{vonStecher2007prl,  YinBlume2015pra}. This reduction of the effective centrifugal barrier is 
reminiscent of Efimov physics, although there is no true Efimov effect in this system even at unitarity, 
i.e. no infinity of bound levels converging to zero energy as one finds for three equal mass bosons at 
unitarity\cite{NaidonReview, DIncaoReview}.


Again, 3NFs have only a minor effect on these systems at short distances, without modifying the long range part.  This relative unimportance appears to be a consequence of the greater Pauli repulsion on a system of 3 or more neutrons, which suppresses the probability for more than two neutrons to come close to each other.  This suppression does not occur for a mixed system of up to four protons and neutrons which can all penetrate to much closer inter-particle or hyperradial distances simultaneously. For this reason, our simple adiabatic potential curve analysis is adequate to explain the absence of both bound and resonant states of the 3n and 4n systems.

Key evidence for our conclusions derives from the energy dependent scattering phaseshift $\delta(E)$ in the lowest adiabatic channel representing the 3n to 3n continuum and for the 4n to 4n continuum, shown \textcolor{black}{as the inset} in Fig.2. Note that, while the results shown here have been obtained in the single-channel adiabatic hyperspherical approximation, numerical tests have also been carried out with full coupled-channel calculations of the multichannel scattering matrix and time delay eigenvalues; there we include all diagonal and off-diagonal nonadiabatic couplings $W_{\nu,\nu'}$, and the results agree quantitatively with the adiabatic results presented here. \textcolor{black}{In fact, the phaseshifts and the time--delays changes only by a few percent when including the couplings with other adiabatic functions.} 


\begin{figure}
\label{fig:Fig3}
\includegraphics[width=8.5 cm] {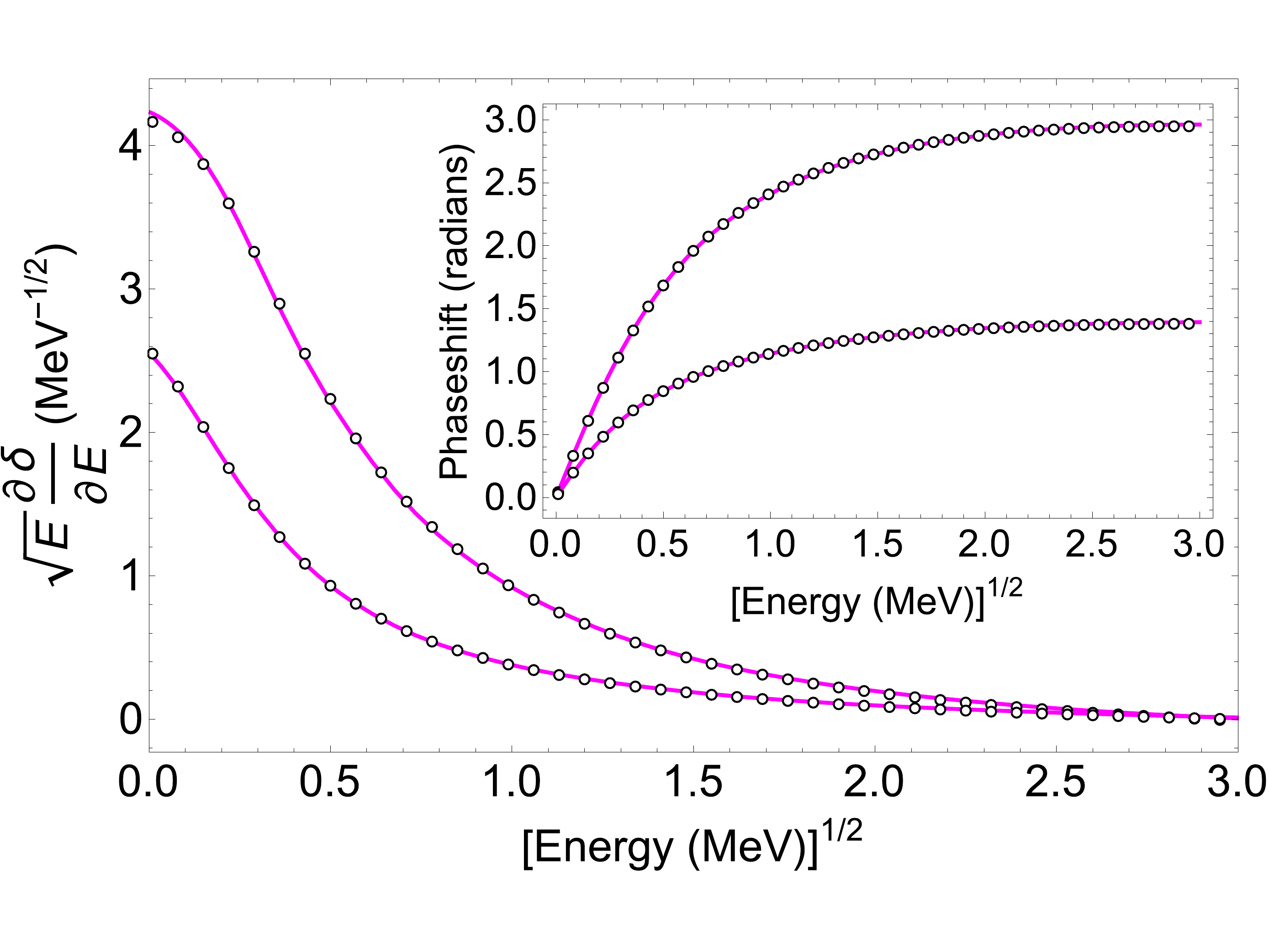}
\caption{Rescaled Wigner-Smith time delays $2 \sqrt{E} d\delta/dE$ for the AV8' interaction (solid curves) and the simple Gaussian interaction (\textcolor{black}{open circles}).  The 4n results are the higher curve, the 3n results the lower.  These are computed in the lowest adiabatic hyperspherical potential energy curve for the 4n $0^+$ symmetry, and for the 3n $\frac{3}{2}^-$ symmetry.  These show no local maximum that would be expected for a low energy resonance.  Note the $E^{-1/2}$ dependence of the Wigner-Smith time delay (or density of states) in the zero energy limit, a consequence of the $\rho^{-3}$ term in the long-range potentials for the 3n and 4n systems. The inset shows the elastic scattering phaseshift versus the square root of the energy.  Both cases show the proportionality to $\sqrt E$ dependence that holds in the zero-energy limit, a consequence of the $\rho^{-3}$ long range potential energy term.  The \textcolor{black}{open circles} that lie almost exactly on top of these curves are computed using the lowest 3n and 4n hyperspherical potential curves based on a simple 2-body Gaussian potential interaction (see text).}
\end{figure}

The Wigner-Smith time delay, defined in general \textcolor{black}{through the scattering matrix $S$} as $Q(E) = i \hbar S dS^\dagger/dE$, which reduces for a single potential curve to $2 \hbar d \delta(E)/dE$, also can be viewed (after division by $2\pi \hbar$) as the density of states enhancement associated with particle interactions. \textcolor{black}{The Wigner--Smith time delay can be viewed as the time difference between an unscattered free--particle wavepacket and the scattered wavepacket off a potential scatterer. A resonance feature would yield a rapid increase in the phase and thus a large time delay, representing the temporary "capture" of an incident wavepacket during the scattering process, resulting in a metastable state \cite{Smith1960PR,TEXIER201616}. The connection between the time delay and the density of states has been shown through a normalization condition of scattering solutions in the contexts of 1D scattering \cite{TEXIER201616}, and multi--channel quantum defect theory \cite{OrangeReview}. A recent application shows a link between the density of states and time delay through a calculation of excess electron distribution of a Rydberg electron near a perturber \cite{giannakeas2020dressed}.}
$Q(E)$ is reported in Fig.2 for the 3n and 4n systems, in each case for both the AV8' and the simple Gaussian interaction; it has been rescaled by $\sqrt{E}$ since the product remains finite at $E\rightarrow 0$. But most critically for our conclusions, the density of states shows no local maximum that would be expected for a low energy resonance in either system.  Both curves do make clear the $E^{-1/2}$ dependence of $Q(E)$ in the zero energy limit, a consequence of the $\rho^{-3}$ term in the long-range potentials for both the 3n and 4n systems.

Consider now the relationship between our present conclusions and some of the alternative theoretical investigations that have been carried out previously for the 3n and 4n systems.  The studies closest to the present spirit, as true scattering theory treatments, are Refs.\cite{ShirokovVary2016prl, Fossez, Deltuva4n2018PL}  There is a strong attraction in the 3n and 4n systems, evidently, but this attraction competes with strong Pauli repulsion.  While the attraction does create a negative $\rho^{-3}$ term in the long range hyperradial potential, it cannot overcome the $\rho^{-2}$ repulsion that is far larger for three or four neutrons than would be the case if two or even one of the particles would be replaced by a proton. \textcolor{black}{It is this dominating repulsion that prevents the lowest 3n and 4n hyperradial potential curves from possessing a local minimum associated with enough attraction to bind these systems or even quasi--bind these systems in the form of a resonance.}  

One fundamental question is the extent to which the 3n and 4n systems fit the pattern of universality that has been well-established for cold fermionic atom systems\cite{petrov2005PRA,petrov2005JPB,Rittenhouse-2011JPB}, especially in the context of the BCS-BEC crossover problem.\cite{regal2007}  We tackle this question by introducing a very simple attractive potential with a single Gaussian for the singlet $nn$ interaction, with a strength and range adjusted to give the correct singlet n-n scattering length and effective range.  Two different choices for the triplet $nn$ interaction have been tested, either neglecting it altogether or setting a Gaussian that reproduces the AV8' p-wave scattering volume and effective range; those two models are indistinguishable on the scale of Figs.1-2.  Results from this simple Gaussian Hamiltonian for the 3n system are shown in the inset of Fig.1(a) as \textcolor{black}{open cirlces} on top of the AV8' results shown as the solid magenta potential curve; remarkably, the results are nearly indistinguishable.  

Finally, we can speculate about the experimental observation of enhanced 4n coincident events in the observation of Kisamori {\it et al.}\cite{Kisamori}.  Even though, in the analysis of that experiment, those enhanced low energy events seemed to indicate existence of a low energy tetraneutron, we speculate that the dramatically enhanced low energy density of states that is evident in our calculations (increasing as $1/{\sqrt E}$) could be the origin of the strong low energy 4n signal.  \textcolor{black}{The fact that an enhanced density of final states produces enhanced cross sections for any process resulting in those final states is familiar from elementary quantum textbook treatments, e.g. of the ``Fermi golden rule''.  We stress that this} enhancement of the 4n density of states is predicted to exist even though no resonance and no bound state exists for the tetraneutron system.

\begin{acknowledgments}
Discussions with Emiko Hiyama and access to her unpublished Gaussian fitted AV8' potential are much appreciated.  The work of MDH and CHG is supported in part by the U.S. National Science Foundation, Grant No.PHY-1912350, and in part by the Purdue Quantum Science and Engineering Institute. The research of AK and MV is supported by the INFN through the National Initiative ``FBS".  
\end{acknowledgments}

\bibliography{tetraneutron_bib1}

\end{document}